\documentclass[twocolumn,showpacs,preprintnumbers,amsmath,amssymb]{revtex4}

\usepackage{graphicx}
\usepackage{dcolumn}
\usepackage{bm}

\def\Xint#1{\mathchoice
   {\XXint\displaystyle\textstyle{#1}}%
   {\XXint\textstyle\scriptstyle{#1}}%
   {\XXint\scriptstyle\scriptscriptstyle{#1}}%
   {\XXint\scriptscriptstyle\scriptscriptstyle{#1}}%
   \!\int}
\def\XXint#1#2#3{{\setbox0=\hbox{$#1{#2#3}{\int}$}
     \vcenter{\hbox{$#2#3$}}\kern-.5\wd0}}

\def\dashint{\Xint-}

\begin{document}

\title{Parametric Statistics of Individual Energy Levels in Random
Hamiltonians}  

\author{I.~E.~Smolyarenko and B.~D.~Simons} 

\affiliation{Cavendish Laboratory, University of Cambridge, Madingley Rd.,
Cambridge CB3 0HE, UK}

\begin{abstract} 
We establish a general framework to explore parametric statistics of
individual energy levels in disordered and chaotic quantum systems of
unitary symmetry. The method is applied to the calculation of the
universal {\em intra-level} parametric velocity correlation function
and the distribution of level shifts under the influence of an
arbitrary external perturbation.
\end{abstract}


\maketitle

Ensembles of random Hamiltonians are used frequently to model
properties of diverse physical systems. Although eigenfunction
statistics can play an important role in some
applications, typically one is interested in the
statistics of the spectra $\{\epsilon_i\}$ of such Hamiltonians. Among
all the possible statistical ensembles~\cite{Zirnbauer} a special role
is played by the three invariant Dyson distributions~\cite{mehta}.
Characterized by the assumption that the distribution is invariant
under, respectively, orthogonal, unitary, and symplectic rotations of
the basis, the random matrix theory (RMT) has proved to be remarkably
successful in modeling physical systems ranging from nuclear
spectra~\cite{porter} and mesoscopic quantum dots~\cite{reviews} to
individual chaotic quantum structures~\cite{BGS}.

An important class of problems arise when individual members $H$
of the statistical ensemble undergo parametric evolution according to the
rule $H \rightarrow H'=H+ X V$, where $V$ is a fixed matrix, and $X$
is the strength of the perturbation. Instead of the random variables
$\epsilon_i$, one is now confronted with {\em random functions}
$\epsilon_i(X)$ of the external parameter $X$. Once cast in terms of the
rescaled variable $x= X\sqrt{\langle (\partial_X\epsilon_i)^2
\rangle}$ (all energies being measured in the units of the mean level 
spacing $\Delta$), it has been argued~\cite{SA} that, for a generic 
perturbation $V$ (see below), the statistical properties of the entire 
random functions $\epsilon_i(x)$ exhibit the same degree of universality 
as that of the parent Hamiltonian $H$. As well as mesoscopic and chaotic 
quantum structures, universality of the random functions $\epsilon_i(x)$ 
finds application to a variety of physical systems including step 
configurations on vicinal surfaces~\cite{teinstein}, non-intersecting 
random walkers in one dimension~\cite{forrester}, and the world lines of 
one-dimensional fermions~\cite{SA}. 

Beginning with the seminal work of Dyson~\cite{Dyson}, the statistical
properties of the random functions $\{\epsilon_i(x)\}$ have been the
subject of numerous investigations \cite{list} (for a review, see,
e.g., Ref.~\cite{efetov}). To date, exact analytic expressions have
been obtained for the distribution of ``local'' (in the parameter
space) properties such as ``level velocities''
$\partial_x\epsilon_i(x)$~\cite{SA,eduardo1,MSS} and ``level
curvatures'' $\partial_x^2 \epsilon_i(x)$~\cite{vonOppen}.  At the
same time, parametric correlations between the sets
$\{\epsilon_i(\bar{x})\}$ and $\{\epsilon_i(\bar{x}+x)\}$ have been
explored~\cite{SA} using field theoretic techniques~\cite{efetov}. As
well as establishing the range of universality, explicit expressions
for the parametric correlator of the density of states (DoS) and the
related level velocity correlation function
$\tilde{c}(\omega,x)=\langle\sum_{ij}
\partial_{\bar{x}}\epsilon_i(\bar{x})\partial_{\bar{x}}\epsilon_j(\bar{x}
+x) \delta(\epsilon-\epsilon_i(\bar{x}))\delta(\epsilon+\omega-
\epsilon_j(\bar{x}+x))\rangle$ have been inferred.

When RMT is applied to many-fermion systems, an important distinction
arises between two classes of correlation functions: Employing the
terminology somewhat loosely, the function $\tilde{c}(\omega,x)$ can
be termed ``grand canonical'' in the sense that the level
$\epsilon_j(\bar{x}+x)$ need not be a parametric ``descendant'' of the
level $\epsilon_i(\bar{x})$. Indeed, the definition of
$\tilde{c}(\omega,x)$ tracks correlations between
$\epsilon_i(\bar{x})$ and a ``descendant'' of {\em any other} level.
However, often one is interested in the parametric evolution of the
Fermi level or a low-lying excitation in systems with a fixed number
of fermions. Inasmuch as such a level can be interpreted as a single
particle level of some effective random many-body Hamiltonian, the
relevant objects are the ``canonical'' correlation functions,
exemplified by the {\em intra-level} velocity correlation function
\begin{equation}
\label{CofX}
c(x)\equiv\left\langle \partial_{\bar{x}}\epsilon_i(\bar{x})
\partial_{\bar{x}}\epsilon_{i}(\bar{x}+x)\right\rangle. 
\end{equation}
A different perspective 
is provided by the distribution of single-level shifts
\begin{equation}
p(\omega,x)\equiv\left\langle\delta\left(\epsilon_i\left(\bar{x}+x\right)-
\epsilon_i\left(\bar{x}\right)-\omega\right)\right\rangle.
\end{equation}
Apart from their intrinsic interest, it has been argued~\cite{reviews}
that the functions $c(x)$ and $p(\omega,x)$ describe parametric
correlations of resonant conductance peaks of quantum dots driven into
the Coulomb Blockade regime and perturbed by a magnetic field or an
external gate potential. Similarly, once $x$ is identified with
magnetic flux, $c(x)$ coincides with the ensemble-averaged correlation
function of single-level persistent currents~\cite{efetov}.

To date, studies of parametric correlations of individual energy levels
have been limited to numerical investigations. Despite its affinity with
$\tilde{c}(\omega,x)$, the intra-level correlation function $c(x)$ (and
its canonical counterparts) belong to a different class of objects. Its
analysis presents technical difficulties which are, in part, similar to
the challenges encountered in the calculation of the level spacing 
distribution in non-parametric random matrix ensembles.
The latter are known to engage DoS correlations which go beyond the
two-point averages presently accessible by field theoretic techniques, 
and lead to expressions in terms of Fredholm determinants with integrable
kernels \cite{mehta}. 

\begin{figure}
\includegraphics[width=3.4in]{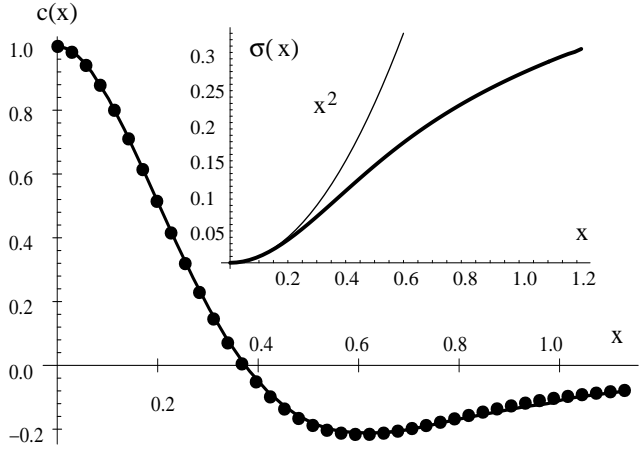}
\caption[]{Level velocity correlation function obtained from 
Eqs.~(\ref{velcorr}), (\ref{z}-\ref{defofd}) (solid line) vs. direct
numerical simulation of large random matrices~\cite{eduardo}
(dots). The width $\sigma\left(x\right)$ of the Gaussian distribution
of level shifts together with the $x^2$ asymptotics at small $x$ is
shown in the inset.}
\label{fig1}
\end{figure}

The aim of the present letter is to formulate a general framework to explore 
parametric correlations of {\em individual} energy levels. In particular,
for the {\em unitary} random matrix ensemble, we will show that 
\begin{eqnarray}
&&c(x)=\int_{-\infty}^{\infty}d\omega\int_0^{2\pi} \frac{d\phi}{ 2\pi}
\frac{1}{z_{+}z_{-}}
\left(-\partial_x^2\right)Z(\omega,x;\phi)\qquad
\label{velcorr}
\\
&&p(\omega,x)=\int_0^{2\pi} \frac{d\phi}{2\pi}\frac{1}{z_{+}z_{-}}
\left(-\partial_\omega^2\right)Z(\omega,x;\phi)
\label{pofx}
\end{eqnarray}
where $z_{\pm}(\phi)=1+e^{\pm i\phi}$ and, adopting the shorthand 
notation $\openone$ to represent the Dirac $\delta$-function,
\begin{equation}
\label{z}
Z\left(\omega,x;\phi\right)=\det_{\left[-1,1\right]}
\left(\openone-\hat{\mathcal K}\left(\phi\right)\right).
\end{equation}
The operator kernel $\hat{\mathcal K}\left(\phi\right)$ 
has matrix elements
\begin{subequations}
\label{kf}
\begin{equation}
\label{k}
{\mathcal K}\left(\lambda,\mu;\phi\right)=\frac{1}{4\pi i}
\sqrt{{\mathcal D}\left(\lambda\right){\mathcal D}\left(\mu\right)}
\frac{F(\lambda;\phi)-F(\mu;\phi)}{\lambda-\mu},
\end{equation}
where 
\begin{equation}
\label{f}
F(\lambda;\phi)=\frac{(z_{+}+z_{-})}{\pi i}
\,\dashint_{-\infty}^{\infty}\!\! d\mu\frac{{\mathcal
D}^{-1}(\mu)}{\mu-\lambda}-(z_{+}-z_{-}) \,{\mathcal D}^{-1}(\lambda)
\end{equation}
\end{subequations}
Here the integral is understood in the sense of the Cauchy principal
value, with variables $\lambda$ and $\mu$ restricted to the interval
$\left[-1,1\right]$, and the dependence on $x$ and $\omega$ is encoded
in the function
\begin{equation}
\label{defofd}
{\mathcal D}\left(\lambda\right)=\exp\left[i\pi\omega\lambda+\pi^2 x^2
\lambda^2/2\right]. 
\end{equation}
It is interesting to note that, after the substitution $x^2\mapsto
-it$, at $\phi=0$, the integral kernel $\hat{\mathcal K}$ coincides
with that arising in the calculation of time-dependent correlation
functions of the one-dimensional interacting Bose gas at zero
temperature~\cite{Izergin}. A comparison of the universal function
$c(x)$ as inferred from Eqs.~(\ref{velcorr}), (\ref{z}-\ref{defofd})
with the results of direct numerical simulation is shown in
Fig.~\ref{fig1}.
%

Before outlining the derivation of these results, several remarks are
in order: (i) The universality of Eqs.~(\ref{kf}) can be inferred from
the universality of Eq.\ (\ref{r}) below \cite{SA}. As such, these
results can be applied to the parametric evolution of spectra which
obey Dyson statistics only ``locally''. (ii) Although we have not
succeeded in obtaining a direct proof, in accordance with the
conjecture made in Ref.~\cite{vallejos}, the distribution of
single-level shifts appears to assume a Gaussian form
$p\left(\omega,x\right)=e^{-\omega^2/2\sigma(x)}/\sqrt{2\pi\sigma(x)}$
{\em at any value} of $x$.  The corresponding width of the Gaussian
can be expressed as
\begin{eqnarray*}
\sigma(x)\equiv\left\langle\left[\epsilon_i(\bar{x}+x)
-\epsilon_i(\bar{x})\right]^2\right\rangle=2\int_0^xdx'(x-x')c(x').
\end{eqnarray*}
At small $x$, where $p(\omega,x)$ can be inferred from the level
velocity distribution~\cite{SA,eduardo1},
$\sigma\left(x\right)\stackrel{x\to 0}{=}x^2$, which reflects
(identifying time with $x^2$) independent ``diffusion'' of individual
levels. In the opposite limit $x\mapsto\infty$, making use of the
known asymptotic dependence~\cite{SA}
$c(x)\stackrel{x\to\infty}{=}-1/\pi^2x^2$ obtained from a perturbative
analysis, one can infer the limiting behavior
$\sigma(x)\stackrel{x\to\infty}{=}\mathrm{const.}+(2/\pi^2)\ln x$.
The resulting strongly sub-diffusive behavior at large $x$ can be
ascribed to the rigidity of the spectrum ``hemming in'' the meandering
levels. (iii) The generating function $Z$ is a particular case of a
more general object $\tilde{Z}_q(J,J';x;\phi)$ which defines
\begin{equation}
P_q(J,J';x)=\int_0^{2\pi}\frac{d\phi }{2\pi}e^{iq\phi}
\tilde{Z}_q(J,J';x;\phi)
\end{equation}
as the probability that the number $n(J)$ of levels
$\{\epsilon_i(\bar{x})\}$ in the (not necessarily contiguous) interval
$J$ and the number $n'(J')$ of levels $\{\epsilon_i(\bar{x}+x)\}$ in
the interval $J'$ differ by exactly $q$. Taking $J$ and $J'$ to be
semi-infinite intervals $J_\epsilon=\left(-\infty,\epsilon\right]$ and
$J_{\epsilon+\omega}=\left(-\infty,\epsilon+\omega\right]$, 
one can derive generalizations of Eqs.~(\ref{velcorr}) and
(\ref{pofx}) which involve correlations between levels
$\epsilon_{i}(\bar{x})$ and $\epsilon_{i+q}(\bar{x}+x)$:
\begin{eqnarray}
\label{cp}
&&c_q(x)\equiv\left\langle \partial_{\bar{x}}\epsilon_{i}(\bar{x})
\partial_{\bar{x}}\epsilon_{i+q}(\bar{x}+x)\right\rangle\\
&& = (-1)^q\int_{-\infty}^{\infty} \!\!\! d\omega\int_0^{2\pi}\!\! 
\frac{d\phi}{2\pi}\frac{e^{iq\phi}}{z_{+}z_{-}}\left(-\partial_x^2\right)
\tilde{Z}_q(J_\epsilon,J_{\epsilon+\omega};x;\phi),
\nonumber\\
\label{pp}
&&p_q(\omega,x)\equiv\left\langle
\delta\left[\epsilon_{i+q}(\bar{x}+x)-\epsilon_{i}(\bar{x})-\omega\right]
\right\rangle\\
&&= (-1)^q\int_0^{2\pi}\frac{d\phi}{2\pi}\frac{e^{iq\phi}}{z_{+}z_{-}}
\left(-\partial_\omega^2\right)
\tilde{Z}_q(J_{\epsilon},J_{\epsilon+\omega};x;\phi).\nonumber
\end{eqnarray}
The exact analytic expression for $\tilde{Z}$ will be given below.
(iv) In some applications, the fixed perturbation $XV$ may be of {\em finite
rank}; i.e., it may possess only a finite number $r$ of non-zero
eigenvalues \cite{SMS}. In such situations Eqs.~(\ref{kf}) retain
their validity providing ${\mathcal D}(\lambda)$ is replaced by
$e^{i\pi\omega\lambda} \det_{\mathrm sc}(1-i\lambda{\mathcal R})$,
where ${\mathcal R}$ is the reactance matrix for scattering off the
potential $XV$ \cite{MSS}, and $\det_{\mathrm sc}$ denotes the
determinant in the space of scattering channels. In this case $x$ may
be identified with any of the variables parametrizing ${\mathcal R}$.
(v) Despite the existence of well-developed analytical tools for the
study of integral kernels with the {\em structure} of (\ref{k})
\cite{Izergin}, it is at present unclear whether these methods can be
generalized to accommodate $\phi$ integration in Eqs.\ (\ref{velcorr})
and (\ref{pofx}).

The analysis of parametric statistics of individual energy levels
relies on a technical device which 
ensures that the level $\epsilon_i(\bar{x}+x)$ is indeed the ``descendant'' 
of $\epsilon_i(\bar{x})$ by demanding that it has the same ordinal number 
as counted from the bottom of the spectrum. Specifically, due to the 
absence of level crossings, the intra-level velocity correlation function
coincides with the conditional average 
\begin{eqnarray}
\label{defofc}
&&c(x)=\int_{-\infty}^{\infty} d\omega \langle\delta_{n(J_{\epsilon}),
n'(J_{\epsilon+\omega})}\nonumber \\
&&\times \sum_{ij} \partial_{\bar{x}}\theta\left[\epsilon-\epsilon_i(\bar{x})
\right] \partial_{\bar{x}}\theta\left[\epsilon+\omega
-\epsilon_j(\bar{x}+x)\right]\rangle,\qquad
\end{eqnarray}
where $\delta_{n,n'}$ denotes the Kronecker $\delta$-symbol, and
$\theta$ is the step-function. The corresponding distribution of level
shifts $p(\omega,x)$ is given by an analogous expression with
$\partial_{\bar{x}}$ replaced by $\partial_{\epsilon}$ (and no
integration over $\omega$). By generalizing the corresponding
non-parametric formula for $P_n(J)$ \cite{footnote}, our starting
point is the general expression for the probability $P_{nn'}(J,J')$ to
find $n$ levels in the interval $J$ of the unperturbed sequence {\em
and} $n'$ levels in the interval $J'$ of the perturbed sequence,
\begin{equation}
\label{pnj}
P_{nn'}(J,J')=\frac{\left(-1\right)^{n+n'}}{n!n'!}\sum_{k=n}^\infty
\sum_{k'=n'}^\infty\frac{\left(-1\right)^{k+k'}r_{kk'}}{\left(k-n\right)!
\left(k'-n'\right)!}.
\end{equation}
Here $r_{kk'}$ represents the multi-point parametric correlation
function of DoS~\cite{SA,SMS} integrated over the interval
$J^k\otimes J^{\prime k'}$ with the corresponding measures $d\mu_J$
and $d\mu_{J'}$. Owing to the determinantal structure of the DoS
correlation function, $r_{kk'}$ can be represented in
the form of a fermionic functional integral
\begin{widetext}
\begin{equation}
\label{r}
r_{kk'}=\det K \int{\mathcal D}\Psi{\mathcal D}\bar{\Psi} \left(\int 
d\mu_J(u)\bar{\xi}(u)\xi(u)\right)^k\left(\int d\mu_{J'}(w)\bar{\eta}(w)\eta
(w)\right)^{k'} e^{\int du\int dw \bar{\Psi}\left(u\right)K^{-1} 
\left(u,w\right) \Psi\left(w\right)}
\end{equation}
\end{widetext}
where $\bar{\Psi} = \left(\bar{\xi},\bar{\eta}\right)$ is a fermionic
doublet. Here
\begin{equation}
\label{kmatrix}
\hat{K}=\left(\begin{array}{cc}\hat{k} & 
\hat{\mathcal D}_0^{-1}\left[\hat{k}-\openone\right]
\\ \hat{\mathcal D}_0 \hat{k} & \hat{k}
\end{array} \right),
\end{equation}
where the matrix elements of the operator sine kernel $\hat{k}$ of the
unitary Dyson ensemble are $k(u-w)=\sin\pi(u-w)/\pi(u-w)$, and
$[\hat{\mathcal D}_0 \hat{k}](u,w)=e^{-(x^2/2)d^2/du^2}k(u-w)$.
Fixing the difference $n'-n=q$ and summing over all $n$, one obtains
the probability $P_q(J,J';x)$ that the numbers of levels in the two
intervals $J$ and $J'$ differ by $q$:
\begin{eqnarray}
\label{pjj}
&&P_q(J,J';x)=\sum_{n=0}^\infty\frac{1}{n!(n+q)!}
\!\!\sum_{\begin{subarray}{c} k=n\\ k'=n+q\end{subarray}}^\infty
\frac{\left(-1\right)^{k+k'+q}r_{kk'}}{\left(k-n\right)!
\left(k'-n-q\right)!}
\nonumber \\
&&=\int_0^{2\pi}\!\!\!\frac{d\phi}{2\pi} e^{iq\phi}
\sum_{\begin{subarray}{c}k=0\\k'=q\end{subarray}}^{\infty}
\frac{\left(-1\right)^{k+k'+q}r_{kk'}}{k!k'!}
z_{+}^kz_{-}^{k'}
\end{eqnarray}
Substituting Eq.~(\ref{r}) into Eq.~(\ref{pjj}), one finds
\begin{eqnarray}
\label{det1}
&&P_q\left(J,J';x\right)=(-1)^q\int_0^{2\pi}\frac{d\phi}{2\pi}e^{iq\phi}
\Bigg\{\det\left[\openone \sigma_0-
\hat{K}\Pi\left(\phi\right)\right] \nonumber \\
&&-\left.\left.
\sum_{k'=0}^{q-1}\frac{\partial_\gamma^{k'}}{k'!}
\det\left[\openone\sigma_0-\hat{K}\Pi_{\gamma}\left(\phi\right)\right]
\right|_{\gamma=0}\right\},
\end{eqnarray}
where 
\begin{eqnarray*}
\Pi\left(\phi\right)=\left(\begin{array}{cc} z_{+} & 0 \\ 0 &  z_{-}
\end{array}\right),\,\,\, \Pi_{\gamma}\left(\phi\right)=
\left(\begin{array}{cc} z_{+} & 0 \\ 0 &
\gamma z_{-}
\end{array}\right),
\end{eqnarray*}
and $\sigma_i$ are the Pauli matrices.  The determinants are
understood as functional determinants on the space of two-component
functions defined on the product interval $J\otimes J'$.  The
expression in curly brackets in Eq.~(\ref{det1}) can be identified as
the generating function $\tilde{Z}$.

In order to apply Eq.\ (\ref{det1}) to the computation of the
intra-level velocity correlation function and the distribution of the
parametric level shifts, one must set $J=J_{\epsilon}$ and
$J'=J_{\epsilon+\omega}$. Setting $q=0$ and using Eq.~(\ref{defofc}),
one obtains Eqs.~(\ref{velcorr}) and (\ref{pofx}), where $Z$ is
identified with the first determinant in Eq.~(\ref{det1}). For $q\ne
0$, a similar procedure leads to Eqs.\ (\ref{cp}) and (\ref{pp}).

The use of semi-infinite intervals to define $n(J_\epsilon)$ and
$n'(J_{\epsilon+\omega})$ is justified only if the support of the
spectrum is finite. The latter condition would be trivially fulfilled
if one were to use, instead of $k(u-w)$, the exact Christoffel-Darboux
kernel of the GUE whose scaling limits interpolate between the sine
kernel inside the Wigner semicircle, and the Airy kernel at its
endpoints. However, in practice, employing such a kernel would present
significant technical difficulties. To circumvent this problem, we use
a regularized kernel
\begin{equation}
\label{kreg}
k_{\delta}\left(u-w\right)=\frac{\sin\pi\left(u-w\right)}
{\pi\left(u-w\right)}
e^{-\left(1/2\right)\delta\left(\left|u\right|
+\left|w\right|\right)},
\end{equation}
where the limit $\delta\rightarrow 0$ is implied in all expressions
involving this kernel. Using Eq.~(\ref{r}) it is easily shown that
$\langle n(J_\infty)\rangle=\langle n'(J_\infty)\rangle=2/\delta$, and
$\langle[n(J_\infty)-n'(J_\infty)]^2\rangle\sim O(\delta)$. Thus,
although the regularization formally violates the level number
conservation, the corresponding error tends to zero in the limit
$\delta\rightarrow 0$. In the following we will suppress the index
$\delta$.

As written, Eq.~(\ref{det1}) involves a matrix oscillating integral
kernel defined on a product of semi-infinite intervals. However, as we
will now show for the case $q=0$, it can be rewritten in the form of
Eqs.~(\ref{kf}) which is (i) more amenable to numerical analysis, and
(ii) makes the integrability of the kernel (in the sense discussed in
Ref. \cite{Izergin}) manifest. Without loss of generality, we can set 
$\epsilon=0$, and shift the variables so as to define the determinant 
on the quadrant $(-\infty,0]\otimes(-\infty,0]$. The corresponding 
shift operator is absorbed into the redefinition $\hat{\mathcal D}_0
\rightarrow\hat{\mathcal D}=e^{\omega d/du}\hat{\mathcal D}_0$. The term 
involving the $\delta$-function in the upper right corner of 
Eq.~(\ref{kmatrix}) can be separated to reveal the dyadic structure of 
the remainder:
\begin{eqnarray*}
\hat{K}=\left(\begin{array}{c}\openone  \\
\hat{\mathcal D}\end{array}\right) \otimes \left( \begin{array}{cc} \hat{k} &
\,\,\,\hat{\mathcal D}^{-1} \hat{k}\end{array}\right)-
\left(\begin{array}{cc}
0 & \hat{\mathcal
D}^{-1}\openone \\ 0& 0\end{array}\right).
\end{eqnarray*}

Now, using the identities $(\openone\sigma_0+z_{-}\hat{\mathcal D}^{-1}
\openone\sigma_{+})^{-1}=(\openone\sigma_0-z_{-}\hat{\mathcal D}^{-1}
\openone\sigma_{+})$, and $\det (\openone\sigma_0+z_{-}\hat{\mathcal
D}^{-1}\openone\sigma_{+})=1$, where $\sigma_{+}=(\sigma_1+i\sigma_2)/2$,
one obtains
\begin{eqnarray*}
&&\det\left[\openone\sigma_0-\hat{K}\Pi\left(\phi\right)\right]\\
&&=\det\left[\openone-\left(z_{+}\hat{k}-z_{+}z_{-}\hat{k}
\left(\hat{\mathcal D}^{-1}
\openone\right)\hat{\mathcal D}+z_{-}\left(\hat{\mathcal
D}^{-1}\hat{k}\right)\hat{\mathcal D}\right)\right].
\end{eqnarray*}
Employing the Fourier representations,
\begin{eqnarray*}
\left\{\begin{array}{c}k(u) \\ \delta(u) \end{array}\right\}
=\frac{1}{2}\int_{-\infty}^{\infty}d\lambda \left\{ \begin{array}{c} 
\theta(1-\lambda)
\theta(1+\lambda) \\ 1\end{array}\right\}  e^{i\lambda\pi u}
\end{eqnarray*}
and making use of Eq.~(\ref{defofd}), one finds
\begin{widetext}
\begin{eqnarray}
&&\left(z_{+}\hat{k}-z_{+}z_{-}\hat{k}\left(\hat{\mathcal D}^{-1}
\openone\right)\hat{\mathcal D}+z_{-}\left(\hat{\mathcal
D}^{-1}\hat{k}\right)\hat{\mathcal D}\right)(u,w)\nonumber\\ 
&&\qquad\qquad =\int_{-1}^1\!\!\!\frac{d\lambda}{2} e^{i\pi \lambda 
\left(w-u\right)}\left\{z_{+}-\frac{z_{+}z_{-}}{2\pi i}\int_{-\infty}^{\infty} 
\!\!\!d\mu \frac{e^{i\pi w\left(\mu-\lambda\right)}{\mathcal
D}^{-1}\left(\mu\right)}{\lambda-\mu-i\delta}\hat{\mathcal
D}+z_{-}{\mathcal D}^{-1}\left(\lambda\right)\hat{\mathcal D}\right\}.
\end{eqnarray}
\end{widetext}
Finally, the cyclic invariance of the determinant and the identity
$z_{+}+z_{-}=z_{+}z_{-}$ are used to perform the integrals in the
$u$-$w$ space, with the resulting kernel in the $\lambda$-$\mu$ space
having the form of Eqs.~(\ref{kf}). Remarkably, taking the limit
$\delta\rightarrow 0$ in the final expressions leads to a {\em
non-singular} kernel defined in terms of the Cauchy principal value
integral~(\ref{f}).

As a final comment, it should be noted that the method of using the
$\phi$ integration to ``count'' the levels in conjunction with the
regularization analogous to (\ref{kreg}) is equally applicable to
other Dyson ensembles. However, at present there exist no analogs
of Eq.\ (\ref{r}) for other ensembles, and thus our consideration is
perforce limited to the unitary case.

\begin{thebibliography}{}
\bibitem{footnote}Eq.\ (\ref{pnj}) is a straightforward generalization
of the non-parametric object $P_n(J)$ (see, e. g., Appendix 7 of
Ref. \cite{mehta}).
\end{thebibliography}
\end{document}